\documentclass[aps,pre,reprint,amsmath,amssymb,superscriptaddress,showpacs,floatfix]{revtex4-1}
\usepackage{graphicx}
\usepackage{dcolumn}
\usepackage{bm}
\usepackage[colorlinks, allcolors=blue]{hyperref}
\usepackage[usenames, dvipsnames]{color}

\begin{document}

\title{Spin-specific heat determination of the ratio of competing first- and second-neighbor exchange interactions in frustrated spin-$\frac{1}{2}$ chains}

\author{Dayasindhu Dey}
\email{dayasindhu.dey@bose.res.in}
\affiliation{S. N. Bose National Centre for Basic Sciences, Block - JD, Sector - III, Salt Lake, Kolkata - 700098, India}

\author{Manoranjan Kumar}
\email{manoranjan.kumar@bose.res.in}
\affiliation{S. N. Bose National Centre for Basic Sciences, Block - JD, Sector - III, Salt Lake, Kolkata - 700098, India}

\author{Si\^an E. Dutton}
\affiliation{Cavendish Laboratory, Department of Physics, University of Cambridge, JJ Thomson Avenue, Cambridge CB3 0HE, United Kingdom}

\author{Robert J. Cava}
\affiliation{Department of Chemistry, Princeton University, Princeton, New Jersey 08544, USA}

\author{Zolt\'an G. Soos}
\email{soos@princeton.edu}
\affiliation{Department of Chemistry, Princeton University, Princeton, New Jersey 08544, USA}

\date{\today}

\begin{abstract}
The magnetic susceptibility $\chi(T)$ of spin-1/2 chains is widely 
used to quantify exchange interactions, even though $\chi(T)$ is 
similar for different combinations of ferromagnetic $J_1$ between 
first neighbors and antiferromagnetic $J_2$ between second neighbors. 
We point out that the spin specific heat $C(T)$ directly determines 
the ratio $\alpha = J_2/|J_1|$ of competing interactions. The $J_1-J_2$ 
model is used to fit the isothermal magnetization $M(T,H)$ and $C(T,H)$ 
of spin-1/2 Cu(II) chains in LiCuSbO$_4$. By fixing $\alpha$, $C(T)$ 
resolves the offsetting $J_1$, $\alpha$ combinations obtained from 
$M(T,H)$ in cuprates with frustrated spin chains.
\end{abstract}

\maketitle

\section{\label{sec:intro}Introduction}
Spin-1/2 chains with isotropic exchange $J_1$, $J_2$ between first and second 
neighbors have been extensively studied both theoretically and experimentally. 
Theoretical interest has focused on the exotic quantum phases of many-spin 
systems with frustrated interactions and variable magnetization in an applied 
field~\cite{sudan2009, furukawa2012, sirker2011}. The ground states are 
analyzed using field theory, density matrix renormalization group (DMRG) 
calculations~\cite{white-prl92, *white-prb93, schollwock2005} and Monte Carlo 
simulations~\cite{sandvik2010}. Crystals that contain edge sharing chains of 
spin-1/2 Cu(II) sites with two bridging oxygen ligands are experimental 
realizations with ferromagnetic ($J_1 < 0$) first neighbor and antiferromagnetic 
($J_2 > 0$) second neighbor exchange~\cite{hase2004}. We refer later to specific 
cuprates. 

The thermodynamics of spin chains, frustrated or not, are obtained by exact 
diagonalization (ED), as pioneered by Bonner and Fisher~\cite{bonner64}, 
or more recently by transfer matrix renormalization group (TMRG) 
calculations~\cite{xiang2006, sirker2010, xiang98}. Isotropic exchange is 
the starting point for detailed magnetic characterization, as recognized 
in linear Heisenberg chains with $J_1$ of either sign. Many kinds 
of extended linear chain compounds are collected in 
Ref.~\onlinecite{*[{}] [{ and references therein.}] miller83}. 
Exchange-coupled chains describe materials with otherwise different spin 
Hamiltonians, and exotic phases or field-induced quantum transitions are typically 
discussed in models with isotropic exchange. 
 
The $J_1-J_2$ model (Eq.~\ref{eq:j1j2} below) with $J_1 < 0$ and $J_2 > 0$ 
has an exact quantum critical point~\cite{hamada88} at 
$\alpha_c = J_2/|J_1| = 1/4$. The ferromagnetic ground state for 
$\alpha < \alpha_c$ switches to a singlet ($S = 0$) for larger $\alpha$ 
The linear Heisenberg antiferromagnet (HAF) has $J_1 > 0$ and $\alpha = 0$ 
in Eq.~\ref{eq:j1j2}. Alternatively, it is the $\alpha \to \infty$ 
limit when Eq.~\ref{eq:j1j2} describes decoupled HAFs on sublattices of 
odd and even-numbered sites. The many exact HAF results~\cite{johnston2000} 
serve as reference for spin chains in general.
 
We model in this paper the thermodynamics~\cite{dutton2012} of the
$J_1 < 0$ chains in LiCuSbO$_4$ 
and show that the spin specific heat directly determines the ratio 
$\alpha = J_2/|J_1|$. The relevant quantities are the magnetization $M(T,H)$ 
and the spin specific heat $C(T,P)$ at temperature $T$ and applied magnetic 
field $H$. In principle, the $T$ and $H$ dependencies of $J_1-J_2$ models are 
fully specified by the exchanges and a scalar $g$ factor, and HAFs illustrate such modeling. 

Multiple quantum phases in frustrated systems are generated by small changes 
of competing interactions. The trade off between $J_1$ and $\alpha$ has 
already been noted in the magnetic susceptibility $\chi (T)$ of the $J_1-J_2$ 
model~\cite{xiang2006, sirker2010}. More negative $J_1$ in the singlet phase 
can be offset by larger $\alpha > 1/4$. By contrast, the spin specific heat 
$C(T)$ is sensitive to $J_1 < 0$ and $\alpha$. The model with $\alpha_c 
< \alpha < 0.40$ has a sharp $C(T)$ peak at low temperature followed by a 
broad maximum, while larger $\alpha$ leads to a single peak~\cite{meisner2006, sirker2010}. 
What has not been appreciated is that $C(T_m)$ at the peak directly specifies $\alpha$
\begin{equation}
C(T_m) = R f(\alpha), \qquad R = k_B N_A.
\label{eq:ctm}
\end{equation}
$R$ is the gas constant. The specific heat is the ideal thermodynamic property 
for quantifying the competition between $J_1 < 0$ and $J_2$.  It has unfortunately 
not been reported in otherwise well studied frustrated spin chains that are 
mentioned in the Discussion. We propose that the specific heat should be 
routinely included when modeling such systems.

An overall modeling of $M(T,H)$ and $C(T,H)$ data with a few parameters is 
challenging and decisive but elementary. It is complementary to the ground 
state properties such as the magnetization $M(0,H)$, exotic quantum phases, 
energy gaps in incommensurate phases or spin correlation functions that are 
obtained by advanced methods. 

\section{\label{sec2} Spin specific heat and magnetization}
We apply standard thermodynamics to the exact energy spectrum of finite systems 
with $2^N$ spin states, just over $1.6 \times 10^7$ for $N = 24$. The $J_1-J_2$ 
model with $|J_1| = 1$ and $S_r = 1/2$ at Cu site $r$ is
\begin{equation}
H(\alpha, h) = -\sum_{r} \vec{S}_{r} \cdot \vec{S}_{r+1} 
+  \alpha \sum_{r} \vec{S}_{r} \cdot \vec{S}_{r+2}  
-h \sum_{r} S_r^z.
\label{eq:j1j2}
\end{equation}
The interaction with the field is $h = g \mu_B H /|J_1|$ where $\mu_B$ is the 
Bohr magneton. We solve at $h = 0$ for $N$ spins and periodic boundary 
conditions. Let $E_{jS}$ be the $j^{\text{th}}$ state in the sector with total 
spin $S \leq N/2$. The Zeeman levels are $-h m_{jS}$ with $m_{jS}$ running 
from $-S$ to $S$. The partition function with $\beta = 1/ k_B T$ of a system 
of $N$ spins is
\begin{equation}
Q_{N}(T,H) =  \sum_{S=0}^{N/2} \sum_{j=1} \sum_{m_{jS} =-S}^{S} \exp 
\left(-\beta (E_{jS} - h m_{jS}) \right).
\label{eq:pf}
\end{equation}
The internal energy is $\langle E_N(T,H) \rangle = - \partial \ln Q_N (T,H)/ 
\partial \beta$. The molar specific heat is
\begin{equation}
C_N(T, H)/R = \left(\beta J_1 \right)^2 \left( \langle E_N (T, H)^2 \rangle - 
\langle E_N (T, H) \rangle^2 \right) / N.
\label{eq:sph}
\end{equation}
The molar magnetization is 
\begin{equation}
M(T,H) = g \mu_{B} \frac{N_{A}}{N} \frac{\partial \ln Q_{N} (T,H)}{\partial\left(\beta h\right)}.
\end{equation}
The molar susceptibility is $\chi(T) = (\partial M(T,H)/\partial H)_0$. We take the reported 
$g = 2.18$ based on electron spin resonance~\cite{grafe2017} of polycrystalline 
LiCuSbO$_4$ and neglect the small, temperature independent diamagnetism or 
van Vleck paramagnetism. $M(T,H)$ is then entirely due to $H(\alpha,h)$. 

The synthesis, structure and thermomagnetic properties of LiCuSbO$_4$ are 
published in Ref~\onlinecite{dutton2012}. $M(T,H)$ and $C(T,H)$ data were 
collected down to $T = 2$ K and 0.1 K, respectively, and up to $\mu_0 H = 16$ T. 
Representative magnetic data, inelastic neutron scattering and limited 
modeling indicated that LiCuSbO$_4$ is a frustrated spin-1/2 chain~\cite{dutton2012}. 
Here we analyze additional isothermal measurements that were collected 
at the same time as the published results on the same sample using the same 
16 T CRYOGENIC Cryogen Free Measurement System (CFMS). The present goal is 
to model quantitatively the entire $M(T,H)$ and $C(T,H)$ data set at $T > 5$~K, 
below which finite-size effects become important.

\begin{figure}
\includegraphics[width=\columnwidth]{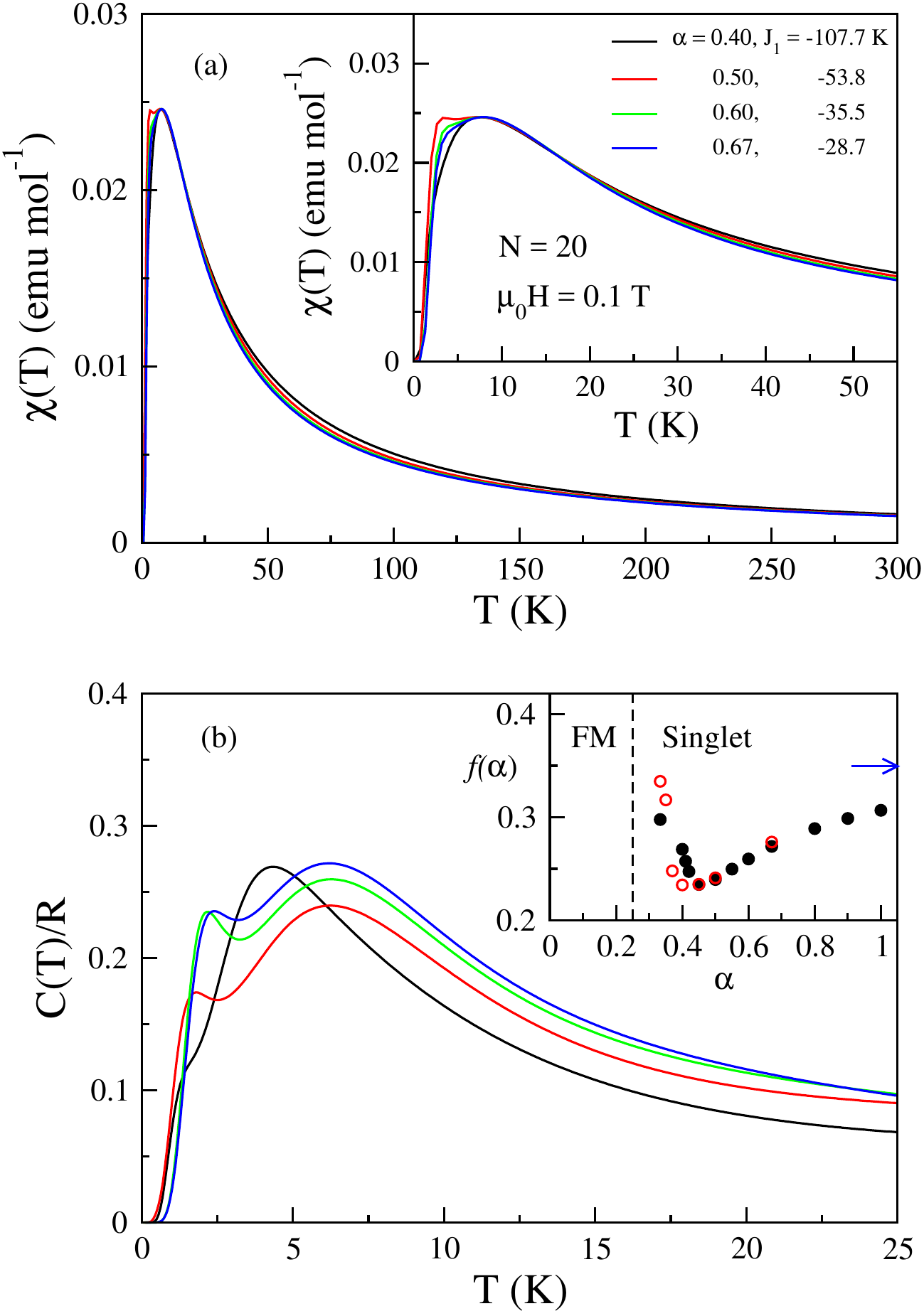}
\caption{\label{fig1}(a) Magnetic susceptibility $\chi (T)$ of $J_1-J_2$ models 
with 20 spins and $\alpha$, $J_1$ chosen to have equal $\chi (T^*)$ at the peak. 
(b) Molar specific heat $C(T)/R$ for the same parameters. The inset shows the 
peak $C(T_m)/R = f(\alpha)$ in the singlet phase; filled points refer to 20 
spins calculations, open point to 24 spins, and the arrow to the limit 
$\alpha \to \infty$.}
\end{figure}

Figure~\ref{fig1}, upper panel, shows 
$\chi(T)$ curves for different parameters that return equal $\chi(T^*)$ at the 
peak. The inset expands the region of the $\chi(T)$ peak. Calculations for 
20 spins with these parameters suffer from finite-size effects below about 
5 K, as demonstrated by comparison with $N = 16$ and 24 results. The size 
dependence is negligible at or above the $\chi(T)$ peak.
Accurate data and careful analysis are needed 
to extract parameters from $\chi(T)$, which is often the first reported 
measurement on prospective spin chains. Reasonable fits are far from unique. 

The zero-field specific heat $C(T)$ in Fig.~\ref{fig1}, lower panel, 
is far more sensitive to the same parameters.
In contrast to $\chi(T)$, there is no trade off: Scaling both exchanges scales 
the peak temperature $T_m$ without changing $C(T_m)$. The inset to the lower 
panel shows $f(\alpha)$ from $\alpha > \alpha_c = 1/4$, where it diverges, 
to $\alpha = 1$. Open and closed circles are exact calculations with $N = 24$ 
and 20 spins, respectively. The open circle at $\alpha = 1/3$ was reported by 
Heidrich-Meisner \emph{et~al.}~\cite{meisner2006} who discussed the numerical 
challenges and used translational symmetry. We also work in k-space with 
periodic boundary conditions in sectors with total $S^z \leq N/2$. The arrow 
marks $f = 0.3497121$ for the HAF~\cite{johnston2000}, the $\alpha \to \infty$ 
limit. The calculated and measured molar specific heat, $C_p \equiv C_v$, of 
the $J_1-J_2$ model with $J_1 < 0$ restricts $\alpha$ to at most two values. Fixing $\alpha$ 
leaves a single exchange, just as in HAFs where magnetic 
data routinely yield the exchange to an accuracy of a few percent. 

The message of Fig.~\ref{fig1} is to start with $C(T,H)$. The zero-field 
peak $C(T_m)$ fixes $\alpha$ of the $J_1-J_2$ model. We then chose $J_1$ to 
fit the susceptibility peak $\chi(T^*)$. Other $M(T,H)$ data could be used 
since the goal is to model all thermodynamics with $J_1$ and $\alpha$.

The measured specific heat is the sum of the spin part, Eq.~\ref{eq:sph}, 
and a lattice contribution, $C_L(T) = a T^3 + b T^5$. The first term is the 
Debye result. Blackman~\cite{blackman41} showed that $T^5$ corrections may 
appear as low as $\Theta_D /50$  where $\Theta_D \sim 200 K$ is the Debye 
temperature. Since $C_L(T)$ is not known separately, we chose a procedure 
that assumes an $H$-independent lattice specific heat. The apparent lattice 
contribution is the difference between the measured specific heat and the 
calculated spin contribution
\begin{equation}
C_{\text{app}}(T, H) = C_{\text{expt}}(T, H) - C_{\text{calc}}(T, H).
\label{eq:capp}
\end{equation}
Perfect agreement with a spin chain collapses the data at all fields to 
$C_L(T) =  a T^3 + b T^5$. Deviations from $C_L(T)$ indicate approximate 
modeling of the spin specific heat.

\begin{figure}
\includegraphics[width=\columnwidth]{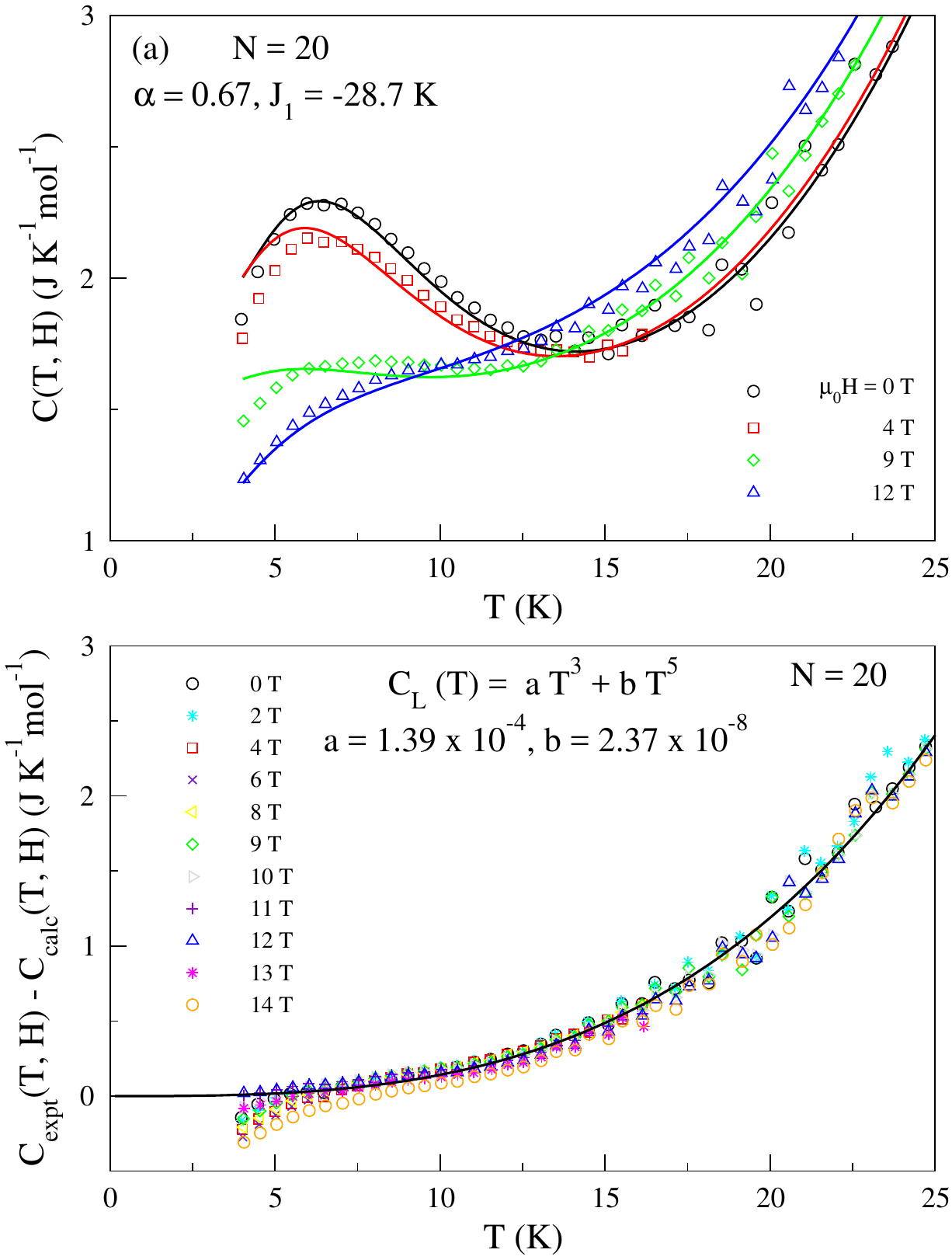}
\caption{\label{fig2}(a) Molar specific heat $C(T,H)$ of LiCuSbO$_4$ at 
$\mu_0 H = 0$, 4, 9 and 12 T. The calculated lines are for $N = 20$ spins in 
Eq.~\ref{eq:j1j2} with $\alpha = 0.67$, $J_1 = -28.7$ K. (b) The lattice 
contribution is $C_L(T) = a T^3 + b T^5$ obtained from Eq.~\ref{eq:capp} 
at the indicated fields $H$.}
\end{figure}

Figure~\ref{fig2}, top panel, shows the experimental $C(T,H)$ of LiCuSbO$_4$ 
at $\mu_0 H = 0$, 4, 9 and 12 T. The field dependence is strong. The calculated 
lines are for 20 spins with $\alpha = 0.67$, $J_1 = -28.7$ K in Eq.~\ref{eq:j1j2}
and $C_L(T)$ obtained from Eq.~\ref{eq:capp}. The lower panel has $N = 20$ results 
at these and other fields. Finite size 
effects appear as expected below 5 K. The apparent lattice contribution 
at higher temperature is almost field independent and follows the Debye law. 

Grafe \emph{et~al.}~\cite{grafe2017} recently discussed LiCuSbO$_4$ by generalizing 
the $J_1-J_2$ model, Eq.~\ref{eq:j1j2}, to have alternating exchanges 
$J_1 (1 \pm \delta)$ along the chain. This is possible in principle since 
there are two Cu atoms per unit cell along the chain and exchange interactions depend 
sensitively on bond lengths and angles~\cite{hatfield83}. LiCuSbO$_4$ has 
chains with equal Cu-Cu separations but slightly different Cu-O bond lengths 
and Cu-O-Cu angles~\cite{dutton2012}. At constant $\alpha$, dimerization 
$\delta$ increases $C(T_m)$. The $C(T,H)$ data in Fig.~\ref{fig2} are almost 
as well fit with $\alpha = 0.55$, $\delta = 0.15$ and $J_1 = -41.1$ K. The 
additional flexibility does improve agreement with experiment in this case. 
We did not search for ($\alpha$, $\delta$) combinations with smaller $\delta$. 
The thermodynamics modeled down to 5 K are compatible with finite $\delta$.
On the other hand, the spin specific heat was overlooked and is clearly 
incompatible with~\cite{grafe2017} $\alpha = 0.28$. We expect that direct 
evaluation of $\alpha$ via $C(T_m)$ will improve the exchange estimates in 
related cuprates with frustrated spin-1/2 chains.

\begin{figure}
\includegraphics[width=\columnwidth]{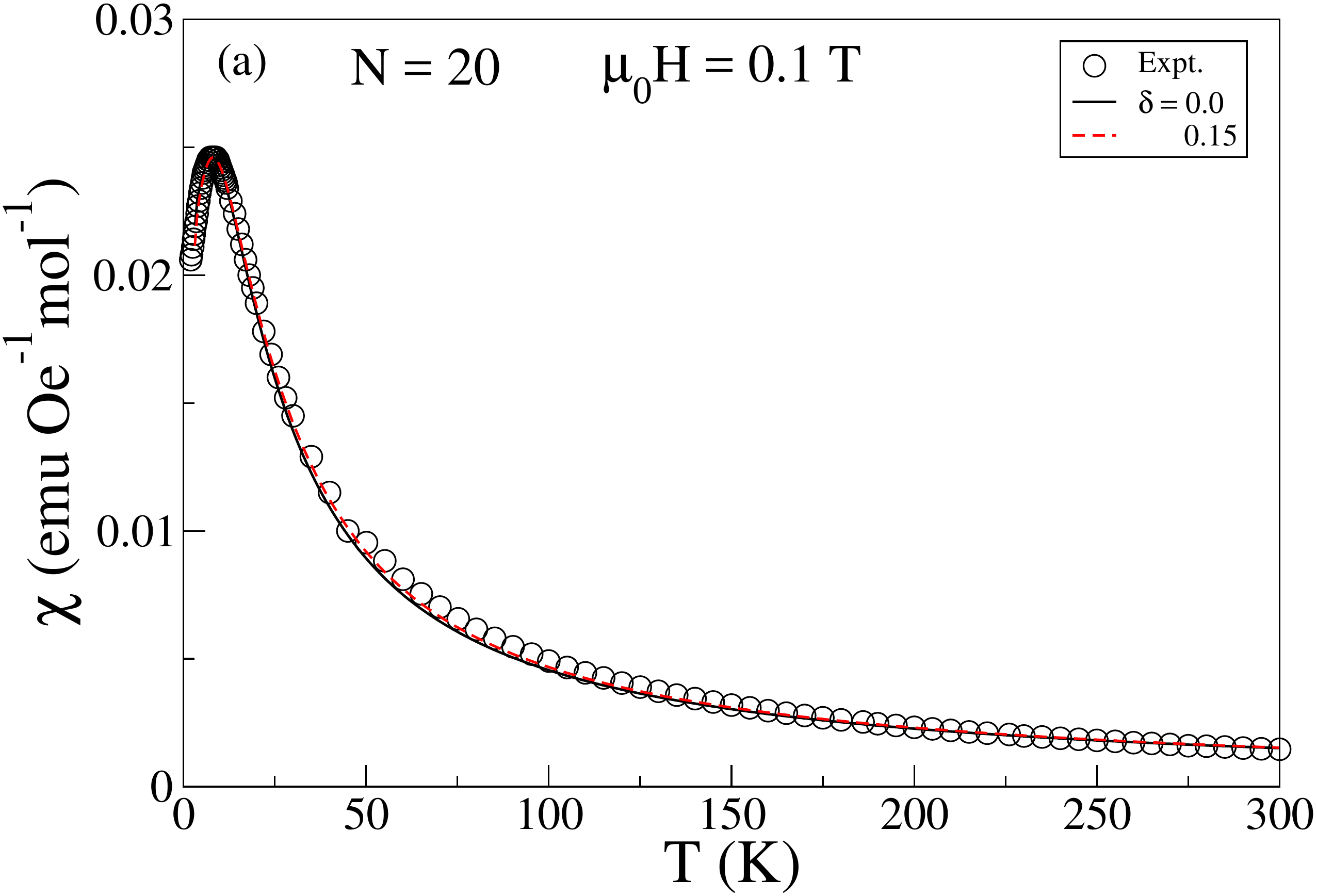}
\includegraphics[width=\columnwidth]{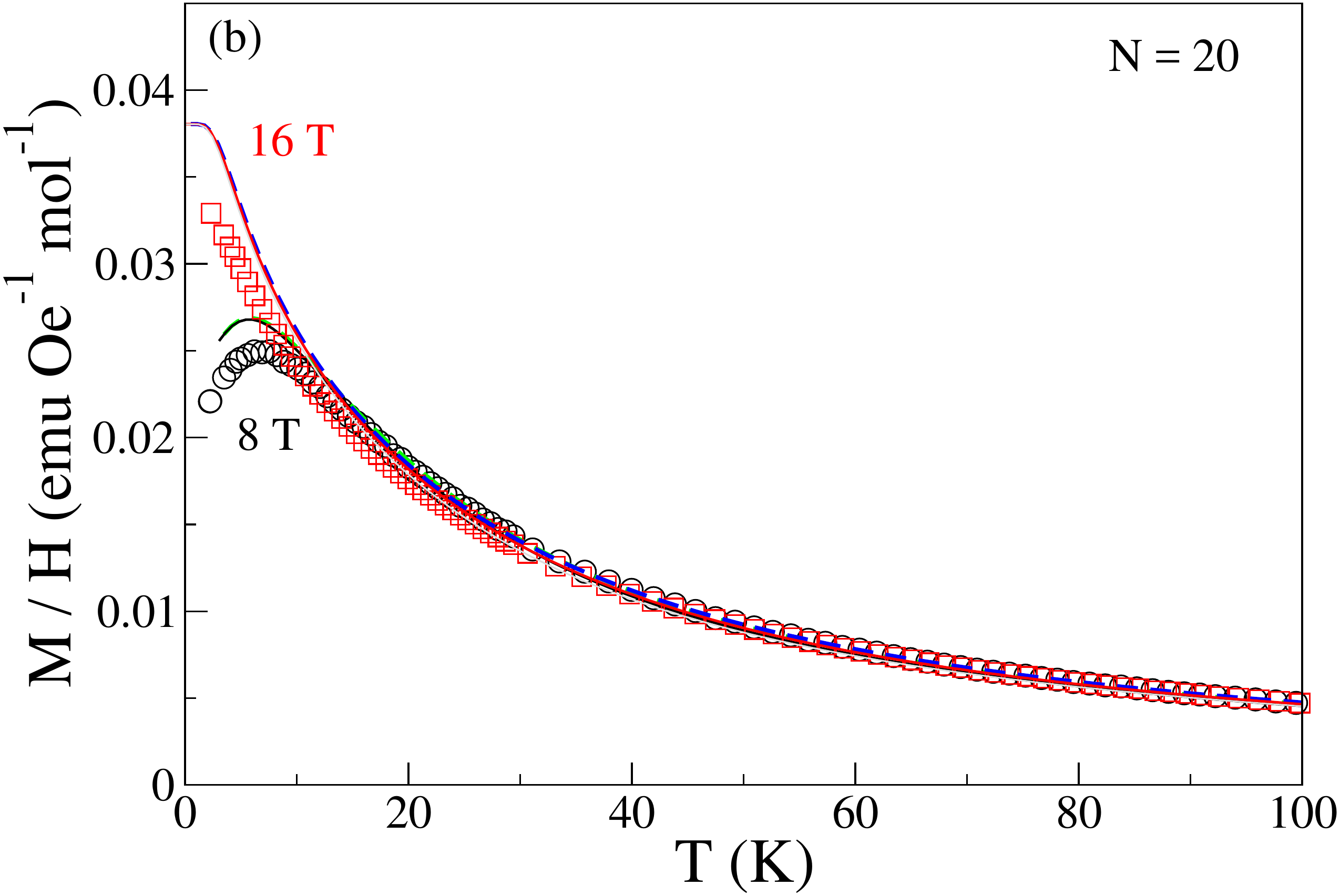}
\caption{\label{fig3}(a) Magnetic susceptibility $\chi(T)$ of LiCuSbO$_4$.
The calculated lines are for 20 spins in Eq.~\ref{eq:j1j2} with (solid line)
$\alpha = 0.67$, $J_1 = -28.7$ K and (dashed line) $\alpha = 0.55$, 
$\delta= 0.15$, $J_1 = -41.1$ K; $J_1$ is chosen to fit the peak $\chi(T^*)$. 
(b) $M(T,H)/H$ vs. $T$ at $\mu_0 H = 8$ and 16 T for the same model parameters.}
\end{figure}

Figure~\ref{fig3}, upper panel, compares the experimental $\chi(T)$ with the 
almost identical calculated susceptibility for $\delta = 0$, $\alpha = 0.67$ 
and $\delta = 0.15$, $\alpha = 0.55$. The lower panel shows the same comparisons 
for $M(T,H)/H$ at $\mu_0 H = 8$ and 16 T with solid and dashed lines for $\delta = 0$ 
and 0.15, respectively. We see again that different parameters return very similar 
magnetic data but distinguishable specific heat. The agreement is good but not perfect. 
The magnetic moment of fully aligned spins is $M = N_A g \mu_B/2$ and gives 
the $M/H = 0.038$ intercept at 16 T. We note that Eq.~\ref{eq:j1j2} has to 
be modified in high fields to tensor rather than scalar $g$ and to include 
deviations from isotropic exchange. We found comparably 
accurate fits for $J_1-J_2$ models with $\alpha$ between 0.40 to 0.67 and 
offsetting $J_1$ chosen as in Fig.~\ref{fig1} to fix $\chi (T^*)$ at the peak. 
Dimerized models with $\delta \sim 0.25$ and $0.4 < \alpha < 0.5$ also fit the 
magnetism and return improved $C(T,H)$ that, however, are less satisfactory 
than shown in Fig.~\ref{fig2}.

\begin{figure}
\includegraphics[width=\columnwidth]{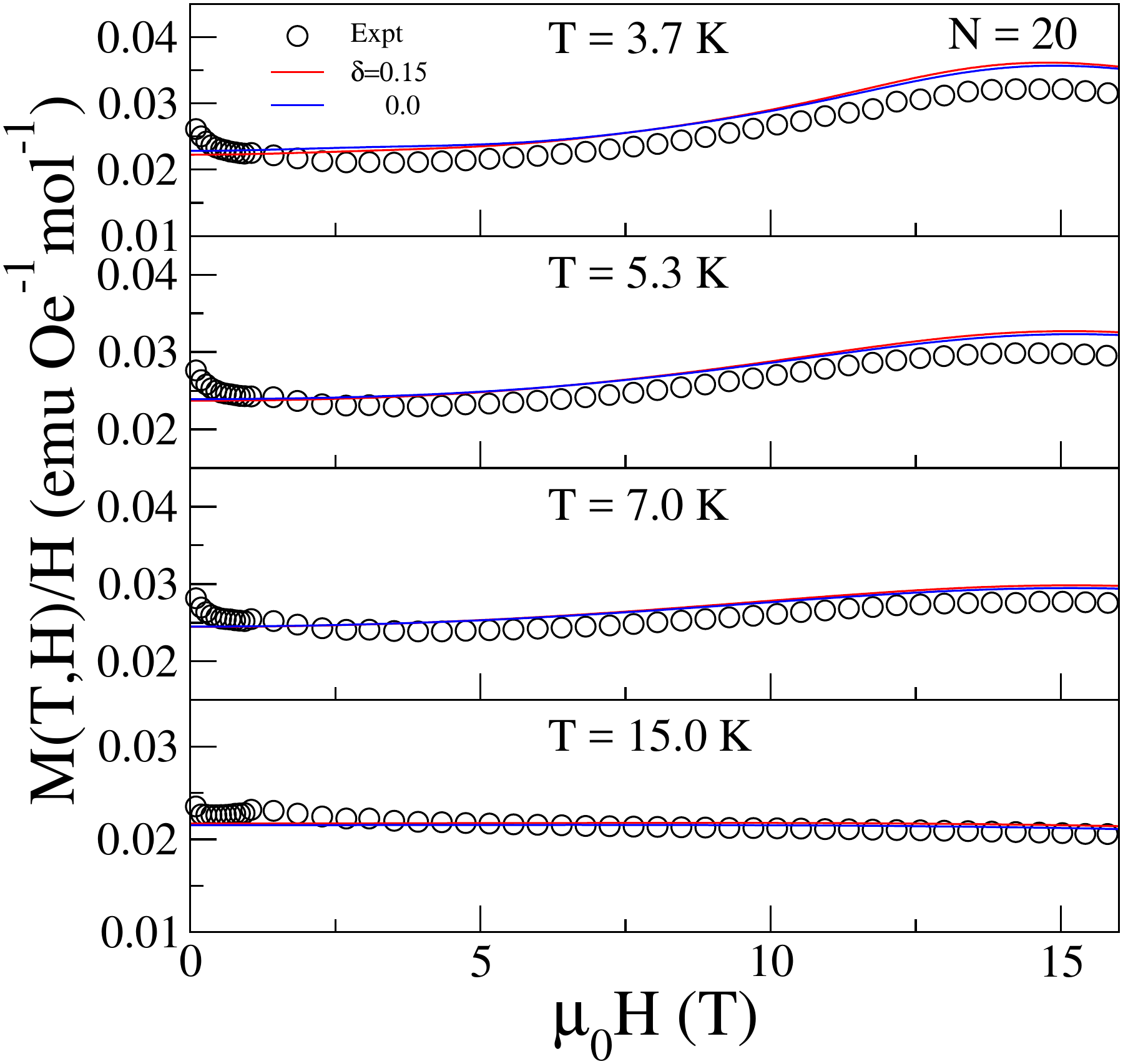}
\caption{\label{fig4} $M(T,H)/H$ vs. $\mu_0 H$ of LiCuSbO$_4$ at the indicated 
temperatures. The calculated blue and red lines are for 20 spins in Eq.~\ref{eq:j1j2} 
with $\alpha = 0.67$, $J_1 = -28.7$ K and $\alpha = 0.55$, $\delta = 0.15$, 
$J_1 = -41.1$ K.}
\end{figure}

Figure~\ref{fig4} shows the field dependence of $M(T,H)/H$ at the indicated 
temperatures. Good fits are obtained at low $H$ or high $T$. DMRG yields the 
ground state magnetization $M(0,H)$ for $N > 100$ spins~\cite{aslam17}. Models 
with isotropic exchange and scalar $g$ have a sharp field-induced transition 
at 0 K to the ferromagnetic state with fully aligned spins. The absolute ground 
state above the saturation field $H_s$ is the Zeeman level $S^z = S = N/2$. 
The calculated $\mu_0 H_s$ are respectively 12.5 and 12.3 T for the $\delta = 0$ and 
0.15 fits. Quite generally, we find $\mu_0 H_s \sim 12$ T for parameters based on 
$\chi(T)$. The measured $dM(T,H)/dH$ at $T = 2$ K shows~\cite{dutton2012} a 
peak centered around 12 T with width of 2 T. More realistically, a $g$-tensor 
yields a range of saturation fields in systems with isotropic exchange. 
Moreover, deviations from isotropic exchange smear out $H_s$ because the total 
spin is then not conserved.   

\section{\label{sec3} Discussion}
All $C(T,H)$ and $M(T,H)$ data for LiCuSbO$_4$ have been analyzed with two 
parameters ($J_1$, $\alpha$) in $J_1-J_2$ models or three parameters 
($J_1$, $\alpha$, $\delta$) in dimerized cases. The field dependence has 
scalar $g = 2.18$ taken from experiment~\cite{grafe2017}. The thermodynamics 
are governed by $H(\alpha, h)$, Eq.~\ref{eq:j1j2}, even though the Hamiltonian 
is known to be approximate and incomplete. It is approximate because 
spin-orbit coupling generates $g$ tensors and deviations from isotropic exchange. 
It is incomplete because the full Hamiltonian has dipole-dipole interactions 
between spins, hyperfine interactions with nuclear spins and various interactions 
between spins in different chains. 

The function $f(\alpha)$ in Eq.~\ref{eq:ctm} and the inset of Fig.~\ref{fig1}(b) 
directly relates the measured maximum $C(T_m)$ of the zero-field specific heat to 
the ratio $\alpha = J_2/|J_1|$. Once $\alpha$ is specified, $J_1$ is found by 
fitting $\chi (T)$ or other magnetic data. The competing interactions of spin-1/2 
chains with $J_1 < 0$ and $J_2 > 0$ are obtained separately. The same parameters 
describe the quantum phases of $J_1-J_2$ models. Ground states properties provide 
other ways to extract $\alpha$ and $J_1$ using field theory or DMRG, but in our 
opinion none is as direct.

We turn to the accuracy of $f(\alpha)$ calculations. Since both ED and TMRG are 
limited to finite temperature, they fail as $\alpha \to \alpha_c$ where 
$T_m/|J_1| \to 0$. Numerical methods return accurate $f(\alpha)$ except close to 
$\alpha_c$. Finite-size effects go roughly as $|J_1|/N$ and are evident at 
$\alpha = 1/3$ where $C(T_m)/R$ increases from 0.298 to 0.334 for $N = 20$ and 24, 
respectively. The corresponding increase at $\alpha = 0.50$ and smaller $|J_1|$
is from 0.239 to 0.245, while $C(T_m)$ at $\alpha = 0.67$ increases by only 
$1.6 \%$ between $N = 20$ and 24. The $N = 20$, $\alpha = 0.67$ fits of $C(T,H)$ 
in Fig.~\ref{fig2} do not change perceptively at $N = 24$ for $T \geq T_m$. 

We conclude that $J_1-J_2$ models with $\alpha_c < \alpha < 1/3$ have $C(T_m)/R > 0.34$. 
That is the range of greatest theoretical interest, close to the quantum critical 
point. To the best of knowledge, however, all reported $C(T_m)/R$ indicate $\alpha > 1/3$. 

We turn briefly to other cuprates with $J_1 < 0$ and $J_2 > 0$. There is no 
indication~\cite{dutton2012} of 3-D ordering in LiCuSbO$_4$ down to 0.1 K, but 
other systems have ordering transitions at lower $T$ than the susceptibility 
peak. Thermodynamic data at finite $T > 5$ K is not sensitive to energy 
differences $\Delta \epsilon \ll k_B T$ that for example differentiate between 
gapped and gapless phases of the $J_1-J_2$ model. 

The measured $C(T)$ of Li$_2$ZrCuO$_2$ has~\cite{drechsler2007} $C(T_m)/R = 0.32$ 
and a fairly sharp peak at $T_m = 6.4$ K that shifts to lower $T_m(H)$ in an 
applied field and is suppressed by 9 T. The $\chi(T^*)$ maximum is 0.037 
emu Oe$^{-1}$mol$^{-1}$, some $50 \%$ higher than the LiCuSbO$_4$ peak in 
Fig.~\ref{fig3}. The inferred ($\alpha$, $J_1$) are~\cite{drechsler2007} 
0.30 and $-273$ K, with $\alpha$ emphasized to be close to $\alpha_c = 1/4$. 
But $\alpha$ is at least 0.35 since $f(1/3)$ returns larger $C(T_m)$ 
and $|J_1|$ is smaller.

The bridging ligands in linarite, PbCuSO$_4$(OH)$_2$, are OH rather than O. 
Single crystals make possible detailed magnetic studies~\cite{wolter2012}, for 
example with $H$ along the principal axes of the $g$ tensor. The inferred 
($\alpha$, $J_1$) from multiple sources are 0.36 and $-100$ K, in line with 
$\mu_0 H_s \sim 7.6$, 8.5 and 10.5 T along the principal axes, but $C(T,H)$ has not 
been reported. 

$M(T,H)$ measurements on Rb$_2$Cu$_2$Mo$_3$O$_{12}$ were originally 
analyzed~\cite{hase2004} as ($\alpha$, $J_1$) with $\alpha = 0.37$ and $J_1 = -138$ K. 
TMRG modeling of $\chi(T)$ is shown in Fig. 8 of Ref.~\onlinecite{xiang2006} for 
the same and related parameters without obtaining a satisfactory fit. We are 
not aware of $C(T,H)$ data. A $J_1-J_2$ model has also been discussed~\cite{masuda2004} 
for neutron diffraction and $\chi(T)$ in LiCu$_2$O$_2$. $C(T)$ was not reported 
and 3D ordering at $\sim 20$ K suggests going beyond a 1D model. 

Banks \emph{et~al.}~\cite{banks2009} performed a comprehensive structural, magnetic 
and computational study of frustrated spin chains in CuCl$_2$. The crystal 
has N\'eel order below $T_c = 23.9$ K. Contributions to the measured $C(T)/T$ 
are estimated~\cite{banks2009} from the lattice ($\sim 80\%$) and from 
overlapping peaks due to the transition and spins. The broad spin peak is at 
$T_m = 35$ K where $C(T_m)/T_m = 0.11$ $\text{J K}^{-1}\text{mol}^{-1}$. These 
numbers return $f(\alpha) = 0.46$ in Eq.~\ref{eq:ctm}, slightly higher than the 
HAF limit (0.35) for a $J_1-J_2$ model with $J_1 < 0$. The lattice 
contribution is obtained indirectly and alternative descriptions are 
mentioned~\cite{banks2009}. So the reported $C(T_m)$ may be consistent 
with $\alpha > 1$ ($f > 0.31$), in line with the overall antiferromagnetism. 

We emphasize in closing that these cuprates are complex systems with diverse 
magnetic, structural, dielectric and other properties. It has been fully recognized 
that the $J_1-J_2$ model is merely the starting point, just as are HAFs for spin 
chains without frustration. In that context, however, the spin specific heat and 
in particular $C(T_m)$ provide a direct evaluation of the ratio $\alpha = J_2/|J_1|$ 
of competing exchange interactions. We expect that $C(T,H)$ measurements will lead 
to more consistent ($J_1$, $\alpha$) parameters for cuprates with frustrated 
spin-1/2 chains.

\begin{acknowledgments}
MK thanks DST for Ramanujan fellowship and computation facility provided under the
DST project SNB/MK/14-15/137. 
\end{acknowledgments}

\bibliography{ref_thermo}
\end{document}